\documentclass[letterpaper,twocolumn,10pt]{article}

\usepackage{usenix}
\usepackage{tikz}
\usetikzlibrary{positioning, calc, arrows.meta}
\usepackage{xspace}
\usepackage{amsmath}
\usepackage{lmodern}
\usepackage{multirow}
\usepackage[capitalise,nameinlink]{cleveref}
\usepackage[T1]{fontenc}
\usetikzlibrary{positioning}
\usepackage{listings}
\usepackage{xcolor}
\usepackage{longtable}
\usepackage{booktabs, array}

\definecolor{codeblue}{RGB}{40,80,160}

\lstdefinestyle{jsonstyle}{
  basicstyle=\ttfamily\footnotesize\color{codeblue},
  breaklines=true,
  breakatwhitespace=false,
  columns=fullflexible,
  keepspaces=true,
  frame=single,
  showstringspaces=false,
  upquote=true
}

\lstdefinestyle{jsonstyleplain}{
  basicstyle=\ttfamily\footnotesize\color{codeblue},
  breaklines=true,
  breakatwhitespace=false,
  columns=fullflexible,
  keepspaces=true,
  showstringspaces=false,
  upquote=true
}

\definecolor{textDark}{HTML}{2B2B2B}
\definecolor{boxFill}{HTML}{E4E7EC}
\definecolor{boxEdge}{HTML}{6B7280}
\definecolor{accentFill}{HTML}{D6E4F0}
\definecolor{accentEdge}{HTML}{3B6BA5}
\definecolor{arrowCol}{HTML}{374151}

\makeatletter
\crefname{algocfline}{line}{lines}
\Crefname{algocfline}{Line}{Lines}
\makeatother
\crefname{algorithm}{Algorithm}{Algorithms}
\Crefname{algorithm}{Algorithm}{Algorithms}
\crefname{section}{\S\!}{\S\S\!}
\Crefname{section}{Section}{Sections}
\crefname{figure}{Figure}{Figures}
\Crefname{figure}{Figure}{Figures}
\crefname{equation}{Equation}{Equations}
\Crefname{equation}{Equation}{Equations}
\crefname{listing}{Listing}{Listings}
\Crefname{listing}{Listing}{Listings}
\crefname{defn}{definition}{definitions}
\crefname{AlgoLine}{line}{lines}
\Crefname{AlgoLine}{Line}{Lines}
\crefname{algocfline}{line}{lines}
\Crefname{algocfline}{Line}{Lines}
\crefalias{algocfline}{line}
\begin{document}

\newcounter{bugsFoundCounter}
\newcounter{bugsAckedCounter}
\newcounter{bugsFixedCounter}

\newcommand{\Found}{Found\stepcounter{bugsFoundCounter}}
\newcommand{\Acked}{Acked\stepcounter{bugsFoundCounter}\stepcounter{bugsAckedCounter}}
\newcommand{\Fixed}{Fixed\stepcounter{bugsFoundCounter}\stepcounter{bugsAckedCounter}\stepcounter{bugsFixedCounter}}

\newcommand{\bugrows}{
\hline
SMTP & Mailpit &
Accepts \texttt{MAIL FROM} before any \texttt{HELO}/\texttt{EHLO}, returning 250 instead of rejecting the command as a bad sequence. &
\Fixed \\
\hline
SMTP & aiosmtpd &
Accepts a syntactically invalid \texttt{MAIL FROM} command without angle brackets and returns 250 instead of rejecting it with a 501 error. &
\Found \\
\hline
SMTP & aiosmtpd &
Accepts invalid recipient syntax \texttt{RCPT TO:Postmaster} (missing angle brackets) and returns 250 instead of rejecting it. &
\Found \\
\hline
SMTP & Mailpit &
Accepts an invalid \texttt{RCPT TO} address containing a malformed source route (missing required colon), returning 250 instead of rejecting the syntax error. &
\Fixed \\
\hline
SMTP & OpenSMTPD &
Rejects a second \texttt{EHLO} issued after \texttt{MAIL FROM} with a 503 error, even though RFC~5321 requires it to reset transaction state. &
\Fixed \\
\hline
SMTP & Mailpit &
Accepts a nested \texttt{MAIL FROM} command during an active transaction (after \texttt{RCPT TO} but before \texttt{DATA}), returning 250 instead of rejecting it. &
\Fixed \\
\hline
SMTP & OpenSMTPD &
Accepts an unqualified local alias in \texttt{MAIL FROM} (e.g., \texttt{<sales>}) after \texttt{EHLO}, returning 250 instead of rejecting it. &
\Acked \\
\hline
SMTP & Stalwart &
Stalwart drops connection after 5 rejected RCPT TO commands &
\Acked \\

\hline
DNS & Technitium &
Returns invalid NS target for apex NS query. &
\Acked \\
\hline
DNS & Bind &
DNS server returns partial TXT RRSet without setting TC bit. &
\Found \\
\hline
DNS & NSD &
DNS server returns partial TXT RRSet without setting TC bit. &
\Found \\
\hline
DNS & Twisted &
DNS server returns partial TXT RRSet without setting TC bit. &
\Found \\
\hline
DNS & GDNSD &
Sets TC for duplicate-identical TXT RRs instead of suppressing duplicates in the RRSet &
\Fixed \\
\hline
DNS & Yadifa &
KEY Record Type (RR Type 25) causes complete zone load failure. &
\Found \\
\hline
DNS & Technitium &
KEY Record Type (RR Type 25) causes complete zone load failure. &
\Found \\
\hline
DNS & CoreDNS &
CoreDNS drops zone records with \texttt{\textbackslash DDD} decimal escape in owner name. &
\Fixed \\
\hline
DNS & Yadifa &
Zone file parser treats \texttt{\textbackslash DDD} decimal-escaped dot as a label separator instead of an intra-label byte. &
\Found \\
\hline
DNS & HickoryDNS &
Authoritative answer returned for name below delegation (zone cut ignored). &
\Acked \\
\hline
DNS & Twisted Names &
Authoritative answer returned for name below delegation (zone cut ignored). &
\Fixed \\

\hline
BGP (Confed) & GoBGP &
Accepts a route whose \texttt{AS\_PATH} contains its own confederation ID/local AS (AS loop not rejected). &
\Fixed \\
\hline
BGP (Confed) & FRR &
Incorrectly rejects routes when its own Member-AS appears in a regular \texttt{AS\_PATH} (non-confederation), treating it as a confederation loop. &
\Found \\
\hline
BGP (Confed) & GoBGP &
GoBGP incorrectly applies AS-loop detection using member-AS instead of confederation ID on eBGP sessions &
\Acked \\
\hline
BGP (Confed) & GoBGP &
Incorrectly establishes an external session and propagates routes between peers in the same Member-AS. &
\Acked \\

\hline
BGP Confed & Batfish &
Does not distinguish AS\_CONFED\_SEQUENCE from AS\_SEQUENCE in BGP RIB. &
\Fixed \\

\hline
BGP (Confed) & GoBGP &
Accepts EBGP session with same confederation ID from non-member peer and installs routes. &
\Found \\

\hline
BGP (Confed) & GoBGP &
Incorrectly accepts confederation-internal session with a non-member peer and installs routes. &
\Found \\

\hline
BGP (Confed) & FRR &
Confed-external peer with remote-as external readvertises route back to sender. &
\Fixed \\

\hline
BGP (Confed) & Batfish &
iBGP route not propagated to peer within same confederation member-AS. &
\Found \\

\hline
BGP (Confed) & FRR &
Rejects route from eBGP peer whose AS equals the confederation identifier. &
\Fixed \\

\hline
BGP (Confed) & FRR &
Does not propagate route from R2 to R3 across inter-confederation eBGP session. &
\Acked \\

\hline
BGP (Confed) & FRR &
Leaks confederation member-AS number in AS\_PATH when advertising to external peer. &
\Fixed \\

\hline
BGP (Confed) & Batfish &
remove-private-as all replace-as not parsed — private AS leaked to external peer &
\Found \\

\hline
BGP (Confed) & Batfish &
BGP session incorrectly established when external peer's AS matches a confederation member-AS. &
\Fixed \\

\hline
HTTP & Caddy &
Serve files when the \texttt{Host} header is present but empty, instead of returning a 400 Bad Request. &
\Fixed \\
\hline
HTTP & Caddy &
Invalid IP-literal values in Host header serves file instead of returning 400. (various examples) &
\Fixed \\
\hline
HTTP & Caddy &
$\%00$ in the request path returns a 500 internal server error&
\Fixed \\
\hline
HTTP & H2O &
$\%00$ in the request path results in a 301 redirect to the same path&
\Fixed \\
\hline
HTTP & Nginx &
Invalid IP-literal values in Host header serves file instead of returning 400. (various examples) &
\Acked \\
\hline
HTTP & H2O &
Serves files when the \texttt{Host} header is entirely missing or present but empty, instead of returning a 400 Bad Request. &
\Found \\
\hline
HTTP & H2O &
Invalid values in Host header serves file instead of returning 400. (various examples) &
\Found \\
\hline

\hline
QUIC & kwik &
Client only offering obselete TLS 1.2 in supported versions passes Handshake instead of failing &
\Fixed \\
\hline
QUIC & quic-go &
Server rejects handshake offering valid TLS 1.3 along with obselete TLS 1.2 in supported versions &
\Found \\
}

\newcommand{\todd}[1]{\textbf{\textcolor{blue}{{TM: #1}}}}
\newcommand{\rathin}[1]{\textbf{\textcolor{orange}{{RS: #1}}}}
\newcommand{\harry}[1]{\textbf{\textcolor{purple}{{HQ: #1}}}}
\newcommand{\george}[1]{\textbf{\textcolor{olive}{{GV: #1}}}}
\newcommand{\siva}[1]{\textbf{\textcolor{brown}{{SK: #1}}}}
\newcommand{\tool}{\textsc{CornerCase}\xspace}
\newcommand{\negParaSpace}{-0.5cm}

\newcolumntype{M}[1]{>{\raggedright\arraybackslash}m{#1}}



\newcommand{\bugsFound}{\arabic{bugsFoundCounter} }
\newcommand{\bugsAcked}{\arabic{bugsAckedCounter} }
\newcommand{\bugsFixed}{\arabic{bugsFixedCounter} }

\newenvironment{soheil}
{\par\color{blue}\noindent\textbf{Soheil: }\ignorespaces}
{\par}

\definecolor{sohblue}{HTML}{0B3D91}
\definecolor{sohdelgray}{HTML}{888888}
\newcommand{\sohnew}[1]{{\color{sohblue}#1}}
\newcommand{\sohdel}[1]{{\color{sohdelgray}\textit{[del]\,#1\,[/del]}}}
\newenvironment{sohblock}
{\par\color{sohblue}\noindent\ignorespaces}
{\par}
\newenvironment{sohnote}
{\par\color{sohblue}\itshape\noindent\textbf{[soheil]\ }\ignorespaces}
{\par}
\newenvironment{sohdelblock}
{\par\color{sohdelgray}\itshape\noindent
 \rule[-0.4ex]{1pt}{1em}\,\textbf{[soheil proposes removing:]}\ \ignorespaces}
{\par\noindent\rule[-0.4ex]{1pt}{1em}\,\textbf{[end removal]}\par}

\date{}

\title{\tool: Automated Extremal Testing of Protocol Implementations}


\newsavebox{\tempbox}
\sbox{\tempbox}{\begin{tabular}{llll}\bugrows\end{tabular}}

\author{
\begin{tabular}{c}
Rathin Singha$^{1}$ \quad Kuan Qian$^{1}$ \quad Srinath Saikrishnan$^{1}$ \\
Tracy Zhao$^{1}$ \quad Soheil Abbasloo$^{2}$ \quad Ryan Beckett$^{2}$ \\
Siva Kesava Reddy Kakarla$^{2}$ \quad Todd Millstein$^{1}$ \quad George Varghese$^{1}$ \\[0.6em]
{\small $^{1}$UCLA \qquad $^{2}$Microsoft Research}
\end{tabular}
}

\maketitle

\begin{abstract}






Many software bugs in network protocol implementations arise near specification boundaries, such as inputs just within or outside allowed ranges, or messages that are valid in isolation but invalid in a given state. From the SSL Heartbleed exploit to TCP Christmas Tree packets, boundary inputs have repeatedly exposed critical weaknesses, yet remain under-tested by existing techniques such as fuzzing and model-based testing. We present \tool, an automated extremal testing approach that systematically targets such boundary behaviors. Our key idea is to decompose test generation into two stages: first, large language models (LLMs) extract explicit validity constraints from protocol specifications (e.g., RFCs) in a structured, section-by-section manner; second, extremal test cases are generated at or near the boundary of each constraint. These tests are executed across multiple implementations, and differential testing identifies inconsistencies. 
We evaluate \tool on widely used implementations of HTTP, DNS, BGP, SMTP, and QUIC, uncovering many previously unknown bugs. 
For example, the HTTP server h2o enters a redirect loop when processing URLs containing encoded null bytes. Overall, we used \tool to identify and file {\bugsFound}anomalies
; to date {\bugsAcked}have been acknowledged as bugs and {\bugsFixed}fixed, with others under active investigation. 
\end{abstract}

\section{Introduction}


As physicists test theories on extreme cases such as infinite mass, software developers routinely test their code on what are colloquially called edge cases that are {\em manually} generated.  Our paper uses AI to {\em automate} the generation of edge cases for network protocol implementations by first using LLMs to extract constraints from protocol specification documents (RFCs). We started this project after realizing that many semantically meaningful edge cases -- such as long DNS names or URLs with invalid characters -- were unlikely to be generated by other automated test generators such as fuzzers~\cite{afl} or testers that use symbolic execution engines~\cite{kakarla2022scale, singha2024messi}.


{\bf Problem:} Elaborating, many real-world software bugs 
appear near the edges of the specification. This problem is exacerbated in network protocol implementations, which have complex input formats with many associated options and constraints. Edge cases include invalid field combinations in messages (e.g., the TCP Christmas Tree attack~\cite{lyon2009nmap}), empty or missing fields, malformed fields, and rare but valid combinations of parameters. Notably, they include messages that violate protocol semantics, such as well-formed protocol messages received in unexpected states. As an example, unexpected handshake messages in TLS (e.g., CLIENT\_HELLO in an invalid state) can lead to crashes or inconsistent state~\cite{de2015protocol}. 
Other examples include the infamous Heartbleed bug~\cite{heartbleed} (CVE-2014-0160) in OpenSSL, where the claimed length was larger than the actual payload.



The attacks above were {\em manually} crafted by hackers, but the rise of LLMs has raised the specter of attackers using AI to {\em automatically} generate such extremal attacks, just as Anthropic uses their Mythos~\cite{anthropic2026mythos} model to identify \emph{memory safety} vulnerabilities. In this paper, we ask: can we use LLMs to automatically generate extremal inputs to harden Internet protocol implementations to (primarily) improve reliability and (secondarily) security, focusing on \emph{message safety} errors -- messages that trigger implementation bugs.

\begin{figure*}[tbh]
    \centering
    \resizebox{\textwidth}{!}{ %
    \begin{tikzpicture}[
    >=Stealth,
    font=\small\sffamily,
    node distance=2mm,
    every node/.style={text=textDark, align=center},
    stepnode/.style={font=\bfseries},
    caption/.style={font=\footnotesize\itshape, text=textDark!75},
    box/.style={draw=boxEdge, fill=boxFill, rounded corners=2pt, inner sep=3pt},
    input box/.style={box, minimum width=2.8cm, minimum height=0.85cm},
    rfc box/.style  ={input box, fill=white, font=\bfseries\small},
    doc box/.style  ={box, minimum width=2.6cm, minimum height=1.2cm},
    res box/.style  ={box, draw=accentEdge, fill=accentFill,
                      minimum width=2.9cm, minimum height=1.5cm},
    flow/.style={->, draw=arrowCol, line width=0.6pt}
]

    \node (user) at (0, 10) {%
        \begin{tikzpicture}[scale=0.28, baseline=-3pt]
            \draw[thick, fill=white, draw=textDark] (0,0) circle (2);
            \fill[textDark] (0,0.4) circle (0.55);
            \fill[textDark] (0,-0.85) ellipse (1.0 and 0.65);
        \end{tikzpicture}%
    };
    \node [stepnode, below=0.15 of user] {User};

    \node (test_in)  [input box] at (3.8, 11.1) {Test Input Format};
    \node (test_out) [input box] at (3.8, 10.0) {Test Output Format};
    \node (rfc)      [rfc box]   at (3.8,  8.9) {RFC};

    \node (ai2) at (7.6, 10) {
        {\includegraphics[width=0.8cm]{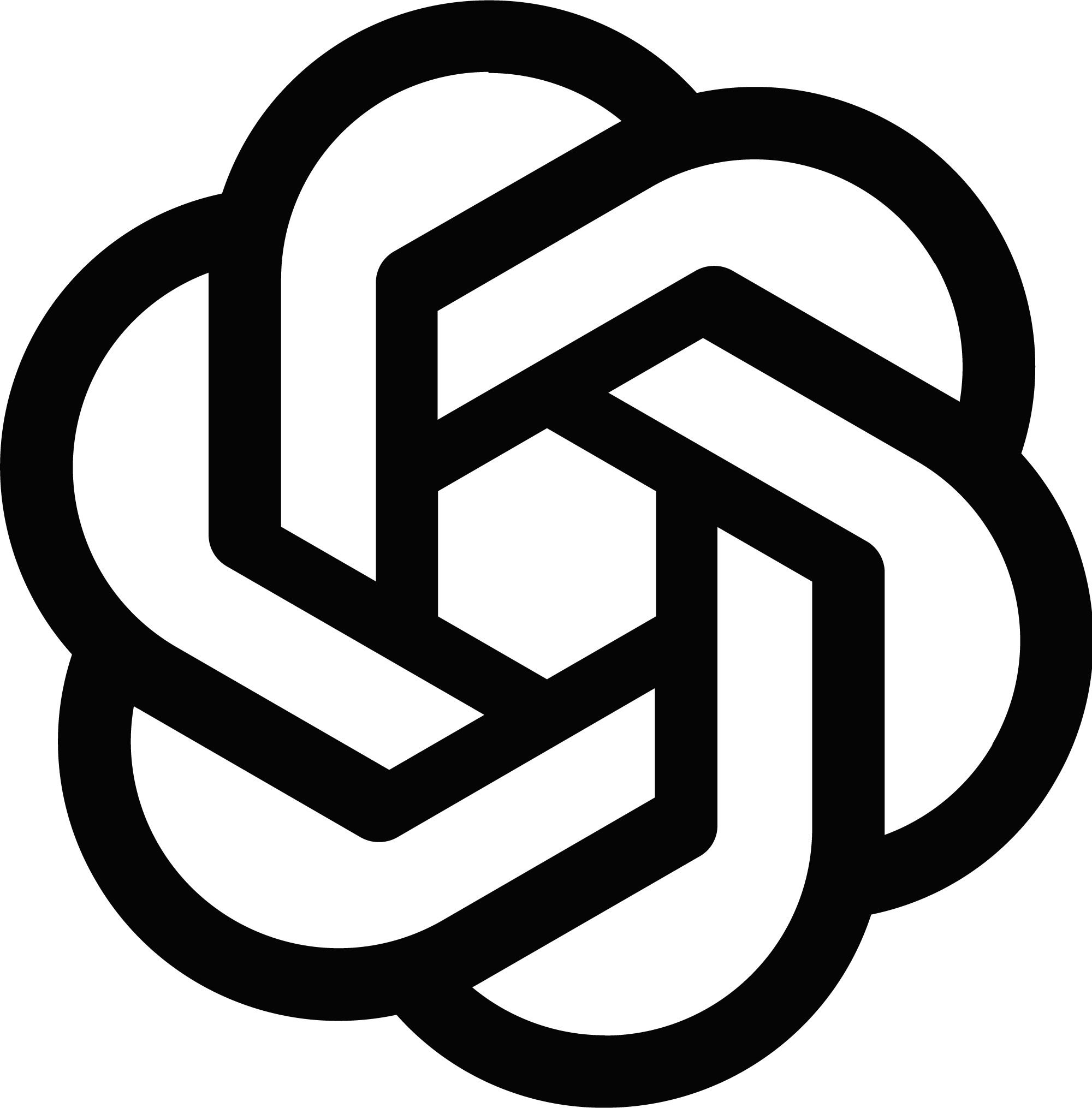}}
    };
    \node [stepnode, below=0.15 of ai2] {1. Constraint \\ Generation};

    \node (constraints) [doc box] at (10.8, 10) {Constraints\\on Inputs};

    \node (ai3) at (14.2, 10) {
        {\includegraphics[width=0.8cm]{images/llm.png}}
    };
    \node [stepnode, below=0.15 of ai3] {2. Test Generation};
    \node [caption, above=0.12 of ai3] {In batches};

    \node (extremal) [doc box] at (17.4, 10) {Extremal\\Test Cases};

    \node (gear) at (11, 8) {%
        \begin{tikzpicture}[scale=0.25, baseline=-3pt]
            \fill[textDark] (0,0) circle (1);
            \foreach \i in {0, 45, ..., 315} {
                \fill[textDark, rotate=\i] (-0.3, 0.8) rectangle (0.3, 1.4);
            }
            \fill[white] (0,0) circle (0.4);
        \end{tikzpicture}%
    };
    \node [caption, above=0.05 of gear] {User Test scripts};
    \node [stepnode, below=0.15 of gear] (step4lbl) {3. Test Execution};


    \node (res_all)  [doc box] at (3.8, 6.0) {Results from \\ Implementations};
    \node (diff)     [doc box] at (9.0, 6.0) {Tests with Diffs};

    \node (ai6) at (11.6, 6.0) {
        {\includegraphics[width=0.8cm]{images/llm.png}}
    };
    \node [stepnode, below=0.22 of ai6] {5. Results\\Analysis};

    \node (res_conf) [res box] at (14.4, 6.0) {Results with\\Confidence Scores};

    \node (python) at (17.2, 6.0) {
        {\includegraphics[width=1cm]{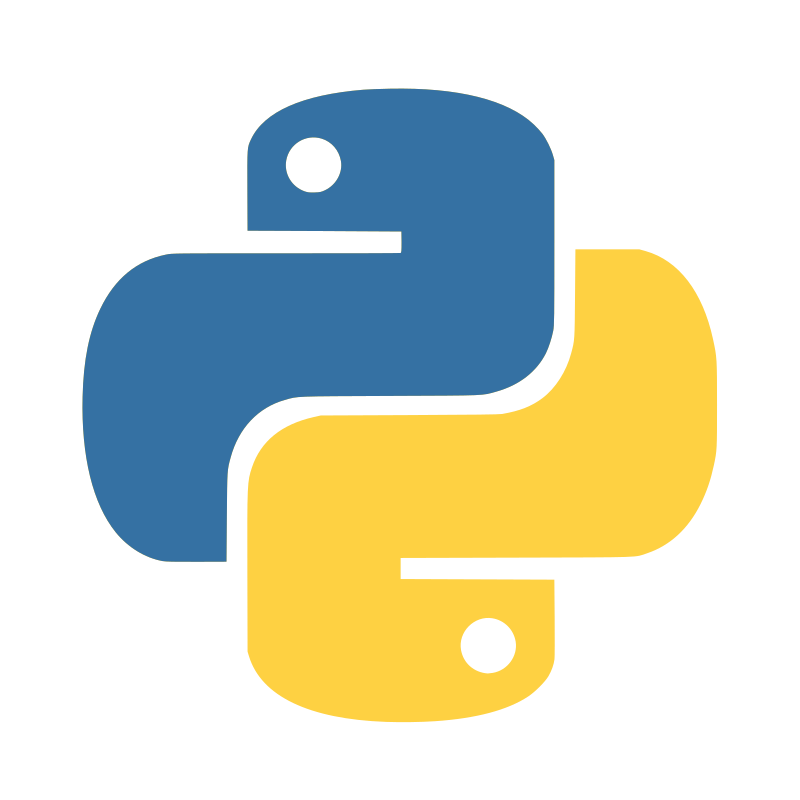}}
    };
    \node [stepnode, below=0.15 of python] {6. Triaging};

    \node (dedup) [res box] at (19.8, 6.0) {Prioritized and \\ Deduplicated \\ anomalies by tag};


    \draw[flow] (user.east) -- ++(0.4,0) |- (test_in.west);
    \draw[flow] (user.east) -- (test_out.west);
    \draw[flow] (user.east) -- ++(0.4,0) |- (rfc.west);

    \coordinate (merge) at (6.1, 10);
    \draw[draw=arrowCol, line width=0.6pt] (test_in.east)  -| (merge);
    \draw[draw=arrowCol, line width=0.6pt] (test_out.east) -- (merge);
    \draw[draw=arrowCol, line width=0.6pt] (rfc.east)      -| (merge);
    \draw[flow] (merge) -- (ai2.west);

    \draw[flow] (ai2.east)          -- (constraints.west);
    \draw[flow] (constraints.east)  -- (ai3.west);
    \draw[flow] (ai3.east)          -- (extremal.west);

    \coordinate (g_in1) at (19.5, 10);
    \coordinate (g_in2) at (19.5, 8);
    \draw[flow] (extremal.east) -- (g_in1) -- (g_in2) -- (gear.east);

    \coordinate (g_out1) at (3.8, 8);
    \draw[flow] (gear.west) -- (g_out1) -- (res_all.north);

    \coordinate (arrow5_mid) at ($(res_all.east)!0.5!(diff.west)$);
    \node (py5) at (arrow5_mid) {{\includegraphics[width=1cm]{images/python.png}}};
    \node [stepnode, below=0.15 of py5] {4. Differential\\Testing};
    \draw[flow] (res_all.east) -- (py5.west);
    \draw[flow] (py5.east)     -- (diff.west);

    \draw[flow] (diff.east)     -- (ai6.west);
    \draw[flow] (ai6.east)      -- (res_conf.west);
    \draw[flow] (res_conf.east) -- (python.west);
    \draw[flow] (python.east)   -- (dedup.west);
    \end{tikzpicture} %
    }

    \caption{The Pipeline for Extremal Testing}
    \label{fig:et_pipeline}
\end{figure*}

{\bf Solution Approach:} We answer this question in the affirmative with \tool, an approach and implementation that uses LLMs in a structured way to generate extremal tests for protocol implementations and analyze testing results (illustrated in~\cref{fig:et_pipeline}). The input to \tool consists of (i) a protocol specification (e.g., an RFC), (ii) a user-defined test input/output format (see Appendix~\cref{app:test-format}), and (iii) a test harness for executing inputs against implementations. The output is a set of extremal test cases together with a ranked set of differential anomalies observed across implementations. This design makes \tool \emph{protocol-agnostic} and {\em implementation-agnostic}. 
In particular, our approach treats implementations as black boxes and does not require source-code access, making it applicable to both open and closed-source systems.


%
%
%

Our approach leverages two properties of many protocols.  First, they have detailed English specifications such as RFCs, which describe validity rules and constraints in natural language, so LLMs provide a natural mechanism for extracting and reasoning about these constraints.
Second, protocols often have multiple independent implementations. This enables us to find bugs via \emph{differential testing}: we can compare behaviors across implementations and flag inconsistencies, which often indicate real bugs or ambiguous parts of the specification.

Methodologically, instead of asking the LLM to generate test cases in one shot, we first use it to extract explicit validity constraints from the specification document, one section at a time, guided by the user-provided test input format to focus on constraints relevant to our test setup. The LLM outputs constraints in a structured form (see Appendix~\cref{app:prompts}) as tuples of constraint ID, section number, and the original RFC sentence which contains the constraint. As part of constraint extraction, we also resolve in-text references to other sections (if any) which are used to define the constraint and add to the extracted tuple.

The key methodological choice behind this decomposition is to use the LLM for \emph{specification understanding} before using it for \emph{test construction}. Asking a model to directly generate protocol tests from a full RFC produces outputs that are broad but shallow: the model misses constraints, under-explores subtle boundaries, and fails to cover the specification systematically. By first extracting explicit constraints and only then generating extremal tests from them, we convert a vague end-to-end generation task into a coverage problem over constraints in the specification. 
We then separately ask the LLM to generate extremal tests (see prompt in~\cref{prompts:test_gen}) for each extracted constraint, ensuring all generated tests conform to the user-provided test format. Each constraint is translated into test cases by modifying only the relevant input fields to produce values just below, at, and just above the specified boundary.

Separating specification understanding from test construction aligns with LLM strengths and avoids their weaknesses. We show using ablation testing (in~\cref{tab:ablation}) that our decomposition produces up to $22\times$ more differential anomalies than one-shot LLM generation.


Each generated test case is executed across multiple implementations using a user-provided test harness that runs inputs against implementations and records outputs in a unified format. We then perform differential analysis to detect inconsistencies in behavior. Finally, we group related differential anomalies using LLM-generated tags that capture the underlying constraint and boundary type. A tag is a label that captures the constraint id and whether the test is just valid or just invalid (see~\cref{sec:methodology_generation}). The LLM also assists in drafting candidate bug reports for manual inspection and editing.

Extremal testing is related to boundary-value analysis (BVA)~\cite{bvareview, sholeh2021black, ramachandran2003testing, zhang2015boundary}, a classical testing technique that targets inputs at the edges of allowed ranges. However, traditional BVA primarily focuses on numeric or range-based constraints and typically relies on manually identified boundaries. By contrast, extremal testing generalizes this idea to a broader class of constraints, extracting syntactic, semantic, and state-based rules from RFC text. A very preliminary version of the conceptual ideas in this paper was described in \cite{singha2025extremaltestingnetworksoftware}.
\noindent {\bf Sample Results and Insights:} Using \tool, we discovered {\bugsFound}bugs across HTTP, DNS, BGP, SMTP, and QUIC implementations. To characterize these, we distinguish between \emph{valid boundary cases} -- inputs that satisfy the specification but lie at the edge of acceptable behavior -- and \emph{invalid boundary cases} -- inputs that are almost well-formed yet violate protocol rules. These boundaries may arise at the syntactic, semantic, or state-machine level. 

For example, the HTTP server h2o~\cite{h2o} responds to requests whose path contains a percent-encoded null byte (e.g., \texttt{GET /$\%00$}) with a \texttt{301} redirect to a semantically equivalent path, inducing a redirect loop that is a potential vector for low-effort denial-of-service attacks (a valid semantic boundary case).

Other examples include malformed \texttt{Host} headers accepted as valid (invalid syntactic boundary cases, one of which is a potential phishing exploit), nested \texttt{MAIL} transactions in SMTP (a state-machine boundary case), incorrect TLS versions accepted during the QUIC handshake (an invalid semantic boundary case), and BGP routes whose \texttt{AS\_PATH} contains the router's own confederation identifier (a valid semantic boundary case). 

These results suggest that many important protocol bugs arise not from arbitrary malformed traffic, but from inputs that sit just inside or just outside specification boundaries, or that are valid in isolation but invalid in context—precisely the regime that extremal testing is designed to expose.

In the process of running \tool, we learned the following lessons that are elaborated on later. 

{\em 1. Focus versus Context:}  Our first prompts missed many constraints until we prompted the LLM to go section-by-section in each RFC to focus the model on a small piece of text.  We did better by adding sections referenced in a section's constraints (more context).


{\em 2. Prompt Templates for General Protocols:} By allowing the user to specify the test format as a JSON file, the framework can be easily extended to new protocols and previously unanticipated features. For example, our HTTP tests improved when we moved from a fixed filesystem -- where the model only generates URI queries -- to a richer format that allows it to specify which files should and should not exist, along with the query.


 {\em 3. Bottleneck Shift:} Our use of LLMs speeds up test and anomaly generation -- so much so that the bottleneck now moves to bug validation and understanding. \tool produces hundreds of discrepancies per protocol, including genuine bugs, artifacts of user misconfiguration, and cases where multiple behaviors are arguably acceptable under the specification. To manage this volume, we use AI for prioritization, tagging, and triage.






    
    


Our paper makes the following contributions:

{\bf 1. Insight:} We show that extremal testing becomes more powerful once LLM use is decomposed into two stages: \emph{constraint extraction} from natural-language specifications, followed by \emph{extremal test generation} from constraints, transforming end-to-end test generation into a coverage problem over constraints in the specification.

{\bf 2. Methodology:} We develop a protocol-agnostic, implementation-agnostic pipeline that uses only RFC text, a user-provided test format, and a black-box execution harness, enabling testing across heterogeneous implementations without source-code access or manual formalization.

{\bf 3. Evidence:} We evaluate the approach on \textbf{38} implementations across \textbf{5} protocols, discovering {\bugsFound}bugs and inconsistencies ({\bugsAcked}acknowledged, {\bugsFixed}fixed). An ablation study shows that each component of the decomposition is essential, with the full pipeline producing up to $22\times$ more anomalies than one-shot LLM generation.



The rest of the paper develops our claims from three angles: the kinds of extremal bugs the method surfaces (in~\cref{sec:motivating}), the structured pipeline that makes those bugs discoverable (in~\cref{sec:methodology}), and empirical evidence that decomposition matters and finds new bugs across protocols (in~\cref{sec:evaluation}). 



\section{Motivating Examples}
\label{sec:motivating}


The following three examples are chosen to illustrate three different ways specification boundaries matter in practice. The HTTP null-byte case lies at a \emph{valid semantic boundary} -- percent-encoding rules permit it syntactically, but its interpretation falls outside what URI-processing components are consistently defined to handle. The BGP confederation example is a \emph{valid semantic boundary}: the message is well-formed, but its validity depends on reading one field (the AS path) in light of another (the router's confederation identity). The SMTP nesting example is a \emph{state-dependent} boundary: each command is individually valid, but the sequence becomes invalid given the protocol state. All three were found using \tool, acknowledged by the relevant developers, and are now fixed.
\subsection{HTTP Null Byte Loops (h2o)}
 RFC 3986 states that a URI with percent-encoded null bytes (\texttt{\%00}) "should be rejected if the application is not expecting to receive raw data" in the component. Historically, null bytes have been used in injection attacks that exploit inconsistencies in low-level string handling. While most modern HTTP servers treat such inputs as malformed and return a \texttt{400 Bad Request} or \texttt{404 Not Found} response in the context of static fileservers, we used \tool to identify two issues with null byte handling in HTTP servers: h2o \cite{h2o} and Caddy \cite{caddy}. The model generated this test twice: from the RFC should statement, and the format rules for percent encoding. 


When a request is issued for a path containing a null byte (e.g., \texttt{GET /\%00}), h2o responds with a \texttt{301 Moved Permanently} redirect whose \texttt{Location} header points to a semantically equivalent path (e.g., \texttt{/\%00/}), by appending a trailing slash. This can induce redirect loops, leading to repeated requests and resource amplification. While individual clients typically bound redirect depth, such behavior can still increase load in aggregate (e.g., from crawlers or automated clients), potentially creating a low-effort denial-of-service vector.

Caddy, by contrast, propagates the malformed path to the filesystem layer, where a null byte in a \texttt{stat} call triggers a system-level \texttt{EINVAL} error that surfaces as a \texttt{500 Internal Server Error}. Rather than rejecting the request as malformed at ingress, Caddy leaks an internal failure mode -- one that may interfere with upstream error handling or monitoring systems that treat 5xx responses as server faults.

\subsection{BGP Confederation Loops (GoBGP)}
BGP requires that routers reject routes whose \texttt{AS\_PATH} contains their own autonomous system (AS) number, in order to prevent routing loops. This rule also applies to BGP confederations. RFC~5065 specifies that if a router receives a route whose \texttt{AS\_PATH} contains its own confederation identifier, it must treat it as an AS loop and reject the route.

Extremal Testing generated a test case where a router receives a route whose \texttt{AS\_PATH} contains its own confederation ID. Unlike the previous example, the received message is syntactically valid, but should be rejected because it violates a semantic constraint.

When executed across implementations, FRR \cite{frr} and Batfish \cite{batfish} correctly rejected the route, treating it as an AS loop. In contrast, GoBGP \cite{gobgp} accepted the route and installed it in the routing table. 
%
Accepting such routes can lead to incorrect behavior and potential loops.

\subsection{SMTP Nesting Failures 
(Mailpit)}
For an example of state-based constraints, SMTP restricts when a new mail transaction may begin. Once a \texttt{MAIL FROM} command has been accepted and one or more \texttt{RCPT TO} commands have been issued, the server expects the client to either send \texttt{DATA} or abort the transaction. Starting a new mail transaction with another \texttt{MAIL FROM} at this point is explicitly forbidden by RFC~5321\textbf{}.

Extremal Testing generated a test case that issues a second \texttt{MAIL FROM} command to a server after it has accepted a recipient, but before \texttt{DATA} is sent. This example is extremal but different from the above two examples: the command is syntactically valid but it is semantically invalid based on the current state of the protocol.

Most SMTP servers that we tested rejected the nested \texttt{MAIL FROM} command with a 503 error or closed the connection. In contrast, Mailpit \cite{mailpit} responded with \texttt{250 2.1.0 Ok}, effectively allowing a new transaction to begin while the previous one was still open. Allowing nested transactions can lead to inconsistent internal state and undefined behavior.

\section{Methodology}
\label{sec:methodology}

Our methodology seeks not just to automate test generation, but to convert natural-language specifications into a structured set of test targets. In that sense, the pipeline is best understood as a way of transforming RFC prose into a coverage object over which boundary-focused testing becomes systematic, traceable, and protocol-agnostic.


\cref{fig:et_pipeline} shows the full pipeline in which we: (1) extract validity constraints from the specification (2) generate many extremal tests that are near the edge of these constraints (3) run these tests across multiple implementations and use differential testing to detect anomalies. We describe each stage below.


\subsection{Input}


Extremal Testing targets systems where (a) there is a written specification (often an RFC), and (b) there are multiple independent implementations that can be tested. These conditions hold for many network protocols.

Our pipeline assumes the following user inputs:

\textbf{Specification document:} An RFC or similar document that describes the protocol and its rules.

\textbf{Input/output test format:} A structured representation of a test case in JSON format, including English descriptions of each test input field and a similar JSON file with test output fields and their description. (See Appendix~\cref{app:test-format} for more details)

\textbf{A tester directory:} A  directory with a main script that can execute test cases in the JSON test input format against each implementation and output results in the JSON test output format.



\subsection{Stage 1: Constraint Generation}
\label{sec:methodology_constraints}

In the first stage of the pipeline, we extract the validity constraints from the specification. The output is a set of explicit constraints that can later be used for systematic test generation.

\vspace{\negParaSpace}
\paragraph{Splitting the specification:} Specifications are often too long to feed into the model at once. We therefore
split the document into sections. We begin with the entire RFC in a single text
document. We then scan the RFC line by line and treat any line matching the
section-header pattern \verb|^\s*(\d+(?:\.\d+)*)\.\s+(.+)$|. as starting a new subsection, where the first capture group is the section
number (for example, 3, 3.1, or 3.1.2) and the second is the title.





For each such header with section number \(k\), we collect all lines from that header up to (but not including) the next header and save the resulting text in a file named \texttt{section\_k.txt}. Here, \(k\) is the section number with dots replaced by underscores (for example, 3.1 becomes
\texttt{section\_3\_1.txt}). We then feed these
\texttt{section\_k.txt} to the LLM one at a time, along with the test format given by user and a prompt to extract all constraints in the text of the section. The LLM returns a list of constraints for each section, which are appended to a global constraint list.

\vspace{\negParaSpace}
\paragraph{Constraint definition:} We define a \emph{constraint} as any sentence in the specification that restricts how protocol inputs may be formed, ordered, or combined -- that is, a boundary between acceptable and unacceptable input that an implementation is expected to enforce.  Concretely, the LLM prompt (see Appendix~\cref{app:prompts}) directs the model to find sentences that describe: syntax rules, allowed or disallowed values, length or size limits, character set restrictions, relationships between multiple inputs, ordering or state rules that can be represented as test inputs etc.

The prompt specifies that constraints are generally RFC statements that use normative language (MUST, MUST NOT, SHOULD, SHOULD NOT), but also includes non-normative sentences that clearly describe a testable rule. Crucially, the LLM is instructed to return each constraint sentence \emph{exactly as written} in the RFC---we do not rewrite, normalize, or formalize constraints at extraction time. Each constraint is recorded as a tuple \verb+["<section_number>", "<constraint sentence>"]+. This avoids introducing interpretation errors and allows later stages to trace each test case directly back to the source.

The test input format plays a key role here: it is included in the prompt so that the LLM can infer which inputs are controllable in the testing framework, and thereby select only those RFC sentences that describe constraints on those inputs. This naturally filters out specification rules that, while valid constraints, cannot be exercised by the test setup (see also the discussion of test-format filtering below).

Based on our extraction process, constraints typically fall into the following categories, illustrated here with concrete examples from our extracted constraint sets:

    
\textbf{1. Range constraints:} numeric bounds on values.
    For example, SMTP reply codes must begin with a three-digit numeric code (RFC~5321, \S2.4), and DNS limits each label to between 1 and 63 octets (RFC~2181, \S11).

\textbf{2. Size constraints:} minimum or maximum sizes for strings, lists, packets, or fields.
    For example, SMTP limits command line length to 512 octets including the \texttt{<CRLF>} (RFC~5321, \S4.5.3.1.4), and DNS limits a full domain name to 255 octets including separators (RFC~2181, \S11).

\textbf{3. Format constraints:} syntactic rules that inputs must follow.
    For example, URI scheme names must consist of a sequence of characters beginning with a letter and followed by any combination of letters, digits, plus, period, or hyphen (RFC~3986, \S3.1), and HTTP header fields must follow a name-value syntax (RFC~9110).

\textbf{4. Dependency constraints:} relationships between multiple inputs.
    For example, in SMTP, the local-part of a mailbox MUST BE treated as case sensitive while verbs and keywords are not (RFC~5321, \S2.4), and in URIs, a percent-encoded octet must be encoded as a character triplet consisting of \texttt{"\%"} followed by exactly two hexadecimal digits (RFC~3986, \S2.1).

\textbf{5. Presence constraints:} conditions under which a field or command is required or forbidden.
    For example, SMTP requires that a client MUST issue \texttt{HELO} or \texttt{EHLO} before starting a mail transaction (RFC~5321, \S4.1.1.1), and TLS~1.3 requires that clients desiring certificate-based server authentication MUST send the \texttt{signature\_algorithms} extension (RFC~8446, \S4.2.3).

\textbf{6. Enumeration constraints:} inputs that must be chosen from a fixed set of allowed values.
    For example, in TLS~1.3, MD5, SHA-224, and DSA MUST NOT be offered or negotiated (RFC~8446, \S4.2.3), and DNS record types must be valid RR types (RFC~1035).

\textbf{7. Ordering and state constraints:} rules about the order in which inputs or commands may appear.
    For example, SMTP specifies that mail transaction commands MUST be used in the prescribed order (RFC~5321, \S3.3). TLS~1.3 requires that protocol messages MUST be sent in the order defined in Section~4.4.1 (RFC~8446, \S4).

\textbf{8. Cross-field semantic constraints:} rules whose validity depends on interpreting the meaning of values across multiple fields or protocol state, beyond simple syntactic checks.
    For example, a BGP confederation member receiving an \texttt{AS\_PATH} containing an autonomous system matching its own AS Confederation Identifier SHALL treat the path as if it contained its own AS number, i.e., reject the route as a loop (RFC~5065, \S4). This constraint cannot be enforced by syntax checks alone -- it requires comparing the \texttt{AS\_PATH} content against the router's own configuration state.


\vspace{\negParaSpace}
\paragraph{Cross-reference detection.}  We detect cross-references inside each constraint sentence using a similar regex pattern. Specifically, we look for textual references of the form ``Section~4.1'', ``Sections~4.1 and~4.2'', etc., and extract all referenced section numbers from the matched group and augment the model's input with the corresponding RFC sections when generating tests.




\vspace{\negParaSpace}
\paragraph{Mapping constraints to the test format:}
The RFC contains many statements that might be categorized as constraints but may not be relevant for our test setup. When we extract constraints from the RFC sections, the setup summary is provided to pick only those constraints that are testable with our testing setup -- e.g. A constraint from BGP RFC 4271 that says,  "KEEPALIVE messages MUST NOT be sent more frequently than one per second" is not relevant if our test setup is testing the decision process and choosing the best route.

\vspace{\negParaSpace}
\paragraph{De-duplication and normalization:}
New constraints are added to the global list only if the pair (section number, constraint) was not seen previously.

\subsection{Stage 2: Test Generation}
\label{sec:methodology_generation}

The goal of this stage is 
to systematically construct tests that lie close to the boundary between valid and invalid behavior, as described by each constraint.

\vspace{\negParaSpace}
\paragraph{Constraint-driven generation:}
Each test case is generated from exactly one RFC constraint sentence. Constraints are passed to the test generator verbatim, together with their RFC section number and the text of any cross-referenced sections. Every generated test explicitly records the constraint that it is intended to exercise. This allows each test to be traced directly back to a specific sentence in the specification.

\vspace{\negParaSpace}
\paragraph{Extremal values:}
For each constraint, we ask the LLM to generate multiple extremal test cases. These include: values that barely satisfy the constraint (e.g., minimum allowed length) and values that barely violate the constraint (e.g., one character too long),
    
    
For numeric constraints such as $0 \leq x \leq 255$, the LLM can generate values like $-1$, $0$, $1$, $254$, $255$, and $256$. For size constraints such as $\texttt{len}(s) \leq 1024$, the LLM generates strings with lengths just below, at, and just above the limit. For format and ordering constraints, the model is instructed to generate inputs that are syntactically valid but placed in invalid positions, or inputs that differ from valid ones by a minimal change.

\vspace{\negParaSpace}
\paragraph{Example test cases:}
To illustrate the kinds of tests generated by Extremal Testing, consider the SMTP constraint in RFC~5321 Section~2.3.5, which states that -- "the reserved mailbox name \texttt{postmaster} MUST be accepted in a \texttt{RCPT} command without domain qualifications." 


From this single constraint, our system generates both positive and negative boundary tests. A positive test checks that \texttt{RCPT TO:<postmaster>} is accepted after a valid \texttt{EHLO} and \texttt{MAIL FROM} sequence. In contrast, negative boundary tests modify the input slightly to violate the constraint, such as using an almost-matching name (\texttt{postmaste}) or an invalid form (\texttt{postmaster@}). These inputs differ by only a single character or structural detail, but must be rejected according to the RFC. Such near-boundary cases are difficult to generate using fuzzing or model-based testing.

\vspace{\negParaSpace}
\paragraph{Tag Generation:} 
Along with the test cases, LLM generates a tag for each of them. The tag is a label that combines the constraint identifier with a boundary indicator (positive or negative), such as \texttt{C11\_Positive} or \texttt{C18\_Negative}. Positive means just valid, and Negative means just invalid test cases.

\vspace{\negParaSpace}
\paragraph{Batching constraints:}
Constraints are processed in small batches (5 by default). Each batch is sent to the LLM in a separate call. Batching is necessary in practice: providing too many constraints at once leads to repetitive outputs and poor coverage. Processing small batches encourages the model to focus on each constraint and produce multiple distinct extremal tests.

\vspace{\negParaSpace}
\paragraph{Context provided to the LLM:}
For each batch, the LLM is given (see prompt in~\cref{prompts:test_gen}): the input test case format in JSON, the exact constraint sentences for the current batch, and, when applicable, the full text of any RFC sections referenced by the constraints.

Providing referenced RFC sections helps the model correctly interpret constraints whose meaning depends on other parts of the specification, such as state transitions or exceptional cases.

\vspace{\negParaSpace}
\paragraph{Test format and identifiers:}
All generated tests must conform exactly to the user-provided input test format. The LLM is instructed to populate each field according to the intended extremal test case. This grounding reduces failures due to unrelated parsing errors and increases the likelihood that tests exercise the intended code paths. The LLM outputs only a JSON array of test objects, with no additional text. Each test object includes a \texttt{constraint} field containing the exact RFC sentence it targets. Test identifiers are assigned post-generation to ensure global uniqueness across batches.


\vspace{\negParaSpace}
\paragraph{Coverage objective:}
The objective of test generation is to maximize the coverage on the RFC constraints, not code coverage. For each extracted constraint, we aim to generate tests that explore behavior just below, at, and just above the specified boundary. 

Overall, this stage converts natural-language specification rules into concrete, boundary-focused test inputs in a systematic and reproducible way, while keeping the generation process simple and scalable.

\subsection{Stage 3: Test Execution}
\label{sec:methodology_execute}


After generation, we run the user-provided test script to execute tests on each implementation and produce a results file in the specified output format. In our experiments, these scripts use Docker containers for each protocol implementation. Given a test case $t$ and an implementation $I$, the tester starts or reuses the corresponding container or process, constructs the concrete input message(s) from $t$, sends them to $I$, waits for a response, and records the output in the expected format.



\subsection{Stage 4: Differential Testing}
\label{sec:methodology_diff}

After executing all test cases, we perform differential analysis to identify cases where implementations behave differently. The goal is not to decide which implementation is correct, but to surface test cases that deserve closer inspection.

\vspace{\negParaSpace}
\paragraph{Diffing logic:}
For each test case, we use a Python script on the results file to compare the outputs across all implementations, each of which is a JSON dictionary. If at least two implementations return different dictionaries for the same test case, we add the test case and all responses to a separate file for detected anomalies. 


\vspace{\negParaSpace}
\paragraph{Output of differential analysis:}
All test cases with response disagreement are logged to a separate file. Each entry includes the original test metadata, and the actual responses from all implementations. This allows later stages to inspect, group, and reason about anomalies without losing contextual information.


\subsection{Stage 5: Result Analysis}
\label{sec:methodology_reports}

Differential analysis can surface a large number of test cases that trigger divergent behavior across implementations. Some of these discrepancies are uninteresting, arising from user misconfiguration or cases where multiple behaviors are arguably acceptable under the specification, while others correspond to genuine bugs or specification violations. Our pipeline uses an LLM to analyze these differences and prioritize them, reducing the human effort required for validation and interpretation.

\vspace{\negParaSpace}
\paragraph{LLM-based analysis of differential results:}
We process the output of differential testing in batches and feed it to an LLM for analysis. Each batch contains a small number of test cases (e.g., five), along with the implementation setup information and the full test metadata. For each test case, the LLM produces a short textual analysis explaining the observed disagreement and assigns a confidence score between 0 and 10, indicating how likely the case represents a real bug or meaningful inconsistency.

The output of this stage is a list of test cases analyzed, each augmented with an LLM-generated comment and confidence score. These results are written to an analysis file that preserves all original test information.

\subsection{Stage 6: Triaging}
\label{sec:triaging}

In the next stage of the pipeline we perform triaging to eliminate semantically similar results and prioritize the results so that manual investigation becomes easy. 

\vspace{\negParaSpace}
\paragraph{Prioritization and grouping:} We sort analyzed test cases by confidence score to prioritize those most likely to correspond to real issues. Since multiple test cases may target the same specification constraint, we group results by their \emph{tag} (see~\cref{sec:methodology_generation}) ---  Within each group, test cases are ranked by confidence score. The pipeline outputs all test cases, grouped and ranked, so that the user can inspect as many or as few as needed. In practice, reviewing the highest-confidence case from each group is an effective starting point, but the full set remains available for deeper investigation.





\vspace{\negParaSpace}
\paragraph{Human validation:}
Results are reviewed manually about whether a case represents a true bug before submission to maintainers. The LLM is used only to assist with analysis, prioritization and deduplication. But this pipeline reduces human effort, allowing us to quickly move from hundreds of differential anomalies to a small, manageable set of high-quality, actionable bug reports.




\subsection{Summary}

The core of Extremal Testing is a two-stage decomposition of specification-based test generation:

\textbf{1. Constraint extraction:} Convert natural-language specification rules into explicit, test-format-aligned constraints.

\textbf{2. Extremal Test generation:} For each constraint, systematically generate tests at and around the specified boundary.


This decomposition gives the pipeline two properties that test generation typically lacks. First, a form of \emph{completeness}: by extracting constraints section by section and tracking which constraints have generated tests, we can measure and improve coverage over the specification. Second, \emph{traceability}: every generated test records the exact constraint sentence and RFC section it targets, so any differential anomaly can be traced directly back to the specification text.


\section{Evaluation}
\label{sec:evaluation}


This section evaluates Extremal Testing on multiple real-world protocols and libraries. Our evaluation is designed to answer three questions. 

\textbf{Q1.} Does Extremal Testing find real, previously unknown protocol bugs? 

\textbf{Q2.} Are those bugs qualitatively different from generic parser failures that existing fuzzers already surface? 

\textbf{Q3.} Is the structured decomposition---section-wise processing, constraint extraction, reference expansion---necessary, or would a one-shot LLM suffice? 


The setup is described in~\cref{sec:exp}. Next the results in~\cref{sec:results} address the first two questions; the ablation study in~\cref{sec:ablation} addresses the third question.




\subsection{Experimental Setup}
\label{sec:exp}

\noindent {\bf Model:} Unless otherwise stated, all prompt-based stages of the pipeline---constraint extraction, test generation, and anomaly analysis---use GPT-5 via the OpenAI API. Full prompts are provided in Appendix~\cref{app:prompts}.We evaluated Extremal Testing on the following implementations (details in~\cref{app:impl-matrix})

\textbf{HTTP servers:} Nginx\cite{nginx}, Apache\cite{apache}, Lighttpd\cite{lighttpd}, Caddy\cite{caddy}, and H2O\cite{h2o}.

\textbf{DNS servers:} BIND \cite{bind9}, NSD \cite{nsd}, Knot \cite{knot}, PowerDNS \cite{PowerDNS}, CoreDNS \cite{coredns}, GDNSD \cite{gdnsd}, Technitium \cite{technitium}, HickoryDNS \cite{trustdns}, TwistedNames \cite{twisted}, Yadifa \cite{yadifa}

\textbf{BGP implementations:} FRR \cite{frr}, GoBGP \cite{gobgp}, Batfish \cite{batfish}.

\textbf{SMTP servers:} AioSMTPD \cite{aiosmtpd}, MailPit \cite{mailpit}, Stalwart \cite{stalwart}, SMTPD \cite{smtpd} and OpenSMTPD \cite{opensmtpd}.

\textbf{QUIC servers:} quiche\cite{quiche}, msquic\cite{msquic}, quic-go\cite{quic-go}, ngtcp2\cite{ngtcp2}, mvfst\cite{mvfst}, kwik\cite{kwik}, picoquic\cite{picoquic}, aioquic\cite{aioquic}, neqo\cite{neqo}, nginx\cite{nginx-quic}, chrome\cite{chrome}, lsquic\cite{lsquic}, haproxy\cite{haproxy}, quinn\cite{quinn}, go-x-net\cite{go-x-net-quic}



All implementations were run inside Docker containers to ensure
reproducibility. 

For HTTP, we test URI parsing rules from RFC~3986. Each test specifies
a URI with scheme, authority, path, query, and fragment components,
along with a filesystem layout with directories and symlinks; we
compare the HTTP status code and resolved file path returned by each
server. For DNS, we test resource record set semantics from RFC~2181,
including duplicate suppression, TTL consistency, and caching
precedence. Each test provides a zone file and one or more queries, and we compare the structured response---including answer
records and return codes---across all resolvers. For SMTP, we test
command syntax and sequencing rules from RFC~5321. Each test specifies
a sequence of SMTP commands (e.g., \texttt{EHLO}, \texttt{MAIL FROM},
\texttt{RCPT TO}) and a server state, and we compare the three-digit
response code returned by each server for the final command. For BGP,
we test confederation handling (RFC~5065) and route reflection
(RFC~4456). Each test specifies an AS topology, confederation
membership, and advertised routes, and we compare whether each
implementation installs the route in its RIB and the resulting AS path
at each router. For QUIC, we test the TLS handshake from RFC~8446.
Each test specifies modifications to an otherwise valid ClientHello message,
and we compare whether the handshake succeeds or fails on each implementation 
using QuicInteropRunner \cite{quic-interop-runner} and 
a modified aioquic\cite{aioquic} client.

We define a \emph{differential anomaly} as a test case where at least
one implementation behaves differently from the others in a meaningful
way---for example, a crash, a different error code (e.g., 4xx
vs.\ 5xx), or acceptance vs.\ rejection of the same input.

For each protocol, we ran the full CornerCase pipeline described in~\cref{sec:methodology}.
As part of its output, the tool extracts input constraints from the specification,
generates boundary-focused test cases, executes them across multiple implementations,
identifies differential anomalies, and applies LLM-assisted analysis to score each
anomaly based on the likelihood of a specification violation. The tool also groups
anomalies by the constraint they exercise and selects the highest-confidence instance
per group to reduce redundancy (\cref{sec:triaging}).
From this ranked output, we use a confidence threshold for each protocol (generally ~8),
lowered when the total number of anomalies is small) to select candidates for manual
inspection, and we submit bug reports for anomalies that pass this manual validation step. During manual inspection, some candidates are discarded because they correspond to implementation design choices, configurable behaviors, or cases where the RFC does not strictly mandate a single behavior. In addition, related anomalies are often grouped into a single report. These cases are excluded from the total number of reported bugs.






\subsection{Differential Anomalies}
\label{sec:results}

\cref{tab:constraints} summarizes the number of unique
constraints extracted, extremal tests generated, differential anomalies found and number of candidates after confidence-based prioritization and triaging -- for each protocol. Each constraint produced between 3 and 11 test cases, including values at the boundary, just below it, and just above it. The candidates after the stage of triaging are manually inspected before passing them to the developers.

\begin{table*}[t]
\centering
\begin{tabular}{llrrrrrr}
\hline
\textbf{Protocol} & \textbf{RFC} & \textbf{Constraints} & \textbf{Tests} & \textbf{Anomalies} & \textbf{Prioritized} & \textbf{Triaged} & \\
\hline
SMTP        & 5321 & 177 & 630 & 526 & 84 & 54 &  \\
DNS         & 2181 &  65 & 213 & 156 & 28 & 23 &  \\
BGP (Confederation) & 5065 &  28 &  98 &  64 & 25 & 20 &  \\
HTTP        & 3986 & 239 & 757 & 297 & 75 & 42 &  \\
QUIC        & 8446 & 136 & 328 &  16 &  6 &  5 &  \\
\hline
\end{tabular}
\caption{Unique constraints, extremal tests, anomalies (disagreement among implementations), prioritized candidates (above the confidence threshold), LLM-triaged candidates (based on tag) per protocol.}
\label{tab:constraints}




\end{table*}





\cref{tab:main_results} summarizes {\bugsFound}bugs found across five protocols (after manual inspection). Of these, {\bugsAcked}have been acknowledged and {\bugsFixed}are fixed by maintainers, and the remainder have been submitted and are
awaiting response. Manual inspection confirms these anomalies are not mere implementation differences: they expose dozens of concrete bugs and specification violations across widely-used protocol stacks, including security-relevant issues in HTTP host validation and BGP loop detection (\cref{tab:main_results}).


\noindent 




A few findings stand out for both their impact and protocol-critical nature. In BGP confederations, we found cases where GoBGP accepted routes with an AS loop and also formed sessions with invalid peer relationships, which can affect routing safety and policy isolation in multi-AS deployments; several of these were acknowledged or fixed. In HTTP, multiple servers accepted malformed or missing \texttt{Host} values but still served content, violating RFC host validation rules and creating a request-routing and virtual-host security risk. In QUIC, version-negotiation behavior around TLS 1.2/1.3 combinations exposed interoperability and downgrade-surface issues, while in SMTP we observed state-machine and syntax-validation errors (for example, accepting \texttt{MAIL FROM} in invalid command sequences), which can enable inconsistent mail handling across implementations. Together, these examples show that extremal differential testing surfaces not only parser edge cases, but also high-value semantic bugs in security and correctness-critical protocol logic.

\vspace{\negParaSpace}
\paragraph{Bug Patterns:} The bugs fall into three broad categories. The most common is \emph{missing validation}, where an implementation accepts syntactically invalid input that the specification requires be rejected---for example, aiosmtpd accepting \texttt{MAIL FROM} without angle brackets, or Caddy serving files for requests with malformed or
missing \texttt{Host} headers. The second category is \emph{incorrect state machine transitions}, where commands are accepted in the wrong order: Mailpit accepts \texttt{MAIL FROM} before any \texttt{EHLO} greeting and allows a nested \texttt{MAIL FROM} during an active transaction. The third is \emph{semantic misinterpretation}, where an implementation misapplies a specification rule. For instance, FRR treats a Member-AS appearing in a regular (non-confederation)
\texttt{AS\_PATH} as a confederation loop, incorrectly rejecting valid routes.

\vspace{\negParaSpace}
\paragraph{Protocol Trends:} BGP confederation bugs involve semantic errors in routing logic, particularly in loop detection and session handling. HTTP bugs are often security-relevant, with malformed \texttt{Host} headers being accepted. SMTP issues stem from weak enforcement of syntax and command sequencing rules, DNS shows robustness problems in handling edge-case semantics and malformed inputs, and QUIC reveals inconsistencies in TLS version negotiation.

\begin{table*}[!h]
\centering
\fontsize{7.5}{8}\selectfont
\begin{tabular}{|M{2.2cm}|M{2.5cm}|M{9.5cm}|M{1.5cm}|}
\hline
\textbf{Protocol} & \textbf{Implementation} & \textbf{Bug Description (brief)} & \textbf{Status} \\
\bugrows
\hline
\end{tabular}



\caption{\textbf{Bugs discovered by Extremal Testing} (total 42 bugs found, 26 acked and 18 fixed). Each row is one distinct implementation-level issue
triggered by an extremal test derived from an RFC constraint.
\emph{Fixed}: patched by maintainers. \emph{Acked}: acknowledged
but not yet fixed. \emph{Found}: reported and under developer
investigation. Bugs span missing validation, state-machine
violations, and semantic misinterpretation across all five
protocols we evaluated.}

\label{tab:main_results}

\end{table*}

\subsection{Ablation Study}
\label{sec:ablation}

We perform an ablation study to identify which parts of the pipeline are responsible for turning raw LLM capability into systematic testing power. The headline result is that each layer of structure matters, and removing any one of them materially reduces anomaly yield (\cref{tab:ablation}).

Our system incorporates three key design choices: (1) processing the RFC section by section (\textbf{S}), (2) extracting explicit constraints from the specification before generating tests (\textbf{C}), and (3) including referenced RFC sections when they are cited in the text (\textbf{R}). We evaluate how each of these components affects the number and usefulness of generated tests.

We compare several variants of the pipeline:

\textbf{Base:} The entire RFC is given to the LLM in one prompt, and the model directly generates test cases without section-wise processing, constraint extraction, or reference expansion.

\textbf{S:} The RFC is fed to the LLM section by section, but the model generates test cases directly without first extracting constraints.

\textbf{SC:} The RFC is processed section by section, constraints are extracted from each section, and test cases are then generated from the collected constraints.

\textbf{SR:} The RFC is processed section by section and referenced sections are included when mentioned, but test cases are generated directly without an explicit constraint extraction step.

\textbf{CR:} The entire RFC is provided at once, but the system extracts constraints before generating tests and includes referenced sections when necessary.

\textbf{SCR (Full System):} Our complete pipeline, which processes the RFC section by section, extracts constraints, and includes referenced sections.

For each variant, we generate tests and run them against all the implementations listed in~\cref{sec:exp}. We record the number of generated tests and the number of differential anomalies discovered across implementations. For variants that perform constraint extraction, we also report the number of extracted constraints. The ablation results across five protocols—DNS, BGP, SMTP, HTTP, and QUIC—are summarized in~\cref{tab:ablation}.




The results show that the full pipeline (\textbf{SCR}) consistently produces the largest number of anomalies across all protocols. Each component contributes meaningfully: section-by-section processing (\textbf{S}) alone yields $\approx10\times$ more anomalies than \textbf{Base} for HTTP (13→131) and $\approx22\times$ for BGP (2→45). Adding constraint extraction further multiplies anomalies—for SMTP, \textbf{SCR} finds 526 vs.\ 124 for \textbf{S} alone. Reference expansion (\textbf{R}) has improved DNS results where \textbf{SCR} finds 156 anomalies vs.\ 100 for \textbf{SC}, probably because some DNS constraints are only fully specified via cross-referenced sections.


The ablation suggests that the central challenge is not getting an LLM to produce tests at all, but getting it to cover the specification \emph{systematically}. Full-RFC prompting (Base) is too diffuse for boundary discovery. Section-wise processing (S) restores focus by localizing reasoning to a tractable unit. Constraint extraction (C) converts the RFC from unstructured prose into an explicit, enumerable set of test targets, turning test generation into a coverage problem. Reference expansion (\textbf{R}) sharpens this set when a constraint's meaning genuinely depends on another section. The pipeline works not because it uses \emph{more} prompting, but because each layer imposes additional \emph{structure} on the model's reasoning.

\smallskip
\noindent {\bf The QUIC outlier:} On QUIC, SC (20 anomalies) slightly exceeds SCR (16). We read this not as a contradiction but as evidence for a general tradeoff visible across the table: additional context helps only when it sharpens a constraint, and can hurt when it broadens the prompt without adding directly testable structure. Cross-references in RFC~8446 are heavy on prose commentary that does not translate into new controllable inputs in our test format, so reference expansion dilutes the prompt without adding coverage. This is consistent with the larger theme of the paper: for extra contextual information not to reduce focus, it must be dense with information that pertains to and enriches  constraints.

\begin{table}[!h]
\centering
\small
\begin{tabular}{|l|l|c|c|c|}
\hline
\textbf{Protocol} & \textbf{Variant} & \textbf{\#Cons.} & \textbf{\#Tests} & \textbf{\#Anom.} \\
\hline

\multirow{6}{*}{\begin{tabular}{l} SMTP\end{tabular}}
 & Base & X & 25 & 20 \\
\cline{2-5}
 & S    & X & 234 & 124 \\
\cline{2-5}
 & SC   & 154 & 613 & 504 \\
\cline{2-5}
 & SR   & X & 284 & 168 \\
\cline{2-5}
 & CR   & 41 & 169 & 95 \\
\cline{2-5}
 & SCR   & 177 & 630 & 526 \\
\cline{2-5}
\hline

\multirow{6}{*}{\begin{tabular}{l} DNS\end{tabular}}
 & Base & X & 20 & 11 \\
\cline{2-5}
 & S    & X & 168 & 71 \\
\cline{2-5}
 & SC   & 63 & 209 & 100 \\
\cline{2-5}
 & SR   & X & 162 & 86 \\
\cline{2-5}
 & CR   & 27 & 96 & 66 \\
\cline{2-5}
 & SCR   & 65 & 213 & 156 \\
\cline{2-5}
\hline

\multirow{6}{*}{\begin{tabular}{l} BGP \\ Confed \end{tabular}}
 & Base & X & 12 & 2 \\
\cline{2-5}
 & S    & X & 68 & 45 \\
\cline{2-5}
 & SC   & 23 & 85 & 49 \\
\cline{2-5}
 & SR   & X & 64 & 33 \\
\cline{2-5}
 & CR   & 20 & 78 & 25 \\
\cline{2-5}
 & SCR   & 28 & 98 & 64 \\
\cline{2-5}
\hline

\multirow{6}{*}{\begin{tabular}{l} HTTP\end{tabular}}
 & Base & X & 27 & 13 \\
\cline{2-5}
 & S    & X & 437 & 131 \\
\cline{2-5}
 & SC   & 165 & 660 & 207 \\
\cline{2-5}
 & SR   & X & 430 & 139 \\
\cline{2-5}
 & CR   & 28 & 124 & 47 \\
\cline{2-5}
 & SCR   & 239 & 757 & 297 \\
\cline{2-5}
\hline

\multirow{6}{*}{\begin{tabular}{l} QUIC\end{tabular}}
 & Base & X & 20 & 0 \\
\cline{2-5}
 & S    & X & 230 & 8 \\
\cline{2-5}
 & SC   & 168 & 344 & 20 \\
\cline{2-5}
 & SR   & X & 336 & 10 \\
\cline{2-5}
 & CR   & 21 & 52 & 1 \\
\cline{2-5}
 & SCR   & 136 & 328 & 16 \\
\cline{2-5}
\hline

\end{tabular}




\caption{\textbf{Ablation of the three structural
components of the pipeline}: section-wise RFC processing
(\textbf{S}), explicit constraint extraction (\textbf{C}), and
targeted reference expansion (\textbf{R}). Columns report
extracted constraints (\#Cons., ``X'' when the variant does not
perform extraction), tests generated (\#Tests), and differential
anomalies (\#Anom.). The full \textbf{SCR} pipeline yields the
highest anomaly count on four of five protocols, producing up to
$22\times$ more anomalies than the one-shot \textbf{Base}.}
\label{tab:ablation}
\end{table}

\subsection{Lessons Learned}

Our experience yields three empirical takeaways: 

\textbf{1. Decomposition beats direct generation:} LLMs are substantially more reliable when asked to extract explicit constraints from a localized portion of a specification than when asked to generate end-to-end tests from an entire RFC. Section-wise processing and constraint-first generation are not implementation details; they are the mechanism that turns LLM capability into systematic test coverage (\cref{tab:ablation}: up to $22\times$ anomaly yield vs.\ one-shot generation).

\textbf{2. Bottleneck shifts from generation to understanding:} With automated test generation, the dominant cost becomes interpreting and de-duplicating differential anomalies. Anomaly ranking, tagging, and triaging therefore become first-class parts of the methodology rather than optional engineering conveniences. For SMTP, we reduced 526 raw anomalies to 84 prioritized candidates and 54 triaged classes, which led to an order-of-magnitude reduction in manual inspection (\cref{tab:constraints}).

\textbf{3. A small interface layer suffices to generalize across protocols:} The only protocol-specific inputs are a test format and an execution harness that expose the relevant controllable inputs and outputs. In practice, this interface was sufficient to carry the same pipeline across HTTP, DNS, BGP, SMTP, and QUIC without retraining or redesigning the core method. Modern code-generation tools (e.g., GitHub Copilot) can bootstrap much of this interface from the specification itself, further lowering the per-protocol cost.

\section{Related Work}
\label{sec:relatedwork}


Related work can be classified by 
what each approach relies on to generate tests or vulerabilities: implementations (fuzzing, symbolic execution, LLM-based unit testing, AI-based vulnerability analysis), past outputs (AI-based vulnerability analysis), formal models (model-based testing), or natural-language specifications themselves. Extremal Testing sits in the fourth category.

\vspace{\negParaSpace}
\paragraph{Boundary value analysis (BVA):} BVA~\cite{bvareview, sholeh2021black, ramachandran2003testing, zhang2015boundary} typically focuses only on range constraints. Extremal Testing is far more general; for instance, it includes state-based constraints for messages. Typically, BVA does not use an LLM to generate boundary conditions. A recent exception \cite{LLMBVA} uses an LLM to directly generate boundary value tests from the \emph{code} while we do so from the \emph{specification}. 
Their evaluation is only for code of a few hundred lines, much smaller than the large code bases we handle.
\vspace{\negParaSpace}
\paragraph{Automated testing:}{\em Fuzz testing} is widely used for software testing in general~\cite{fuzzer1, fuzzer2, peachfuzzer, bohme2016coverage, 10.1145/2714565} and specifically for BGP and DNS implementations (e.g., \cite{forescoute, frr-fuzzer, cambus, nmap, honggfuzz}). Although fuzzers are effective at finding parser bugs, random inputs have only a small probability of finding extremal bugs.
{\em Symbolic execution} uses SMT solvers to generate test cases for many execution paths of a program (e.g., \cite{cadar2008klee,godefroid2005dart}). Protocol implementations contain many thousand lines, making symbolic execution infeasible. Extremal testing, unlike symbolic execution, works on software whose source code is unavailable. 
{\em Model-based testing} uses an abstract model of a system to generate tests~\cite{bozic2018formal}. It has been used for DNS \cite{kakarla2022scale} and BGP~\cite{singha2024messi} but generates valid, not extremal inputs.  Eywa~\cite{eywa} uses an LLM to generate models, not tests.


\vspace{\negParaSpace}
\paragraph{LLM driven test generation:} LLM based software testing is a vast area~\cite{LLMTestingSurvey}, a subset of LLM based Software Engineering~\cite{FacebookLLMTesting}.  However, existing work~\cite{LLMTestingSurvey} uses LLMs to improve coverage or to improve mutation/fuzz based testing, not for extremal testing. Our framework generates end-to-end tests on multiple implementations instead of unit tests for individual implementations.  


\vspace{\negParaSpace}
\paragraph{AI based vulnerability analysis:} Recent work uses LLMs for security, including automated penetration testing~\cite{pentestgpt} and code-level vulnerability detection~\cite{zhou2024llm_vuln_detection}. LAPRAD~\cite{rustem} first trains an LLM on prior DNS attacks and then constructs new attacks; by contrast, we generate tests based only on RFCs.


\section{Limitations and Future Work}
\label{sec:limitations}

Extremal testing currently has the following limitations:

{\em Implicit Constraints:}  The constraints defining the Heartbleed~\cite{heartbleed} exploit (payload length does not match length field) and Christmas Tree packets~\cite{de2015protocol} (SYN, FIN, and RST flags should not be all set) are what one might call \emph{implicit constraints}: they are not explicitly stated in the RFC but are strongly implied. It may be
possible in future work to ask an LLM to generate implicit constraints by a new prompting strategy. 

{\em Short interactions:} We currently focus on generating extremal tests that involve a single message or short command sequence. Extending to protocol behaviors that involve long multi-step workflows, timing dependencies, or complex state machines is interesting future work.


{\em Single RFCs:} Constraint extraction is currently based on single RFCs. Protocol behavior can depend on multiple RFCs referencing each other.
-- e.g., RFC 9113 (HTTP 2) uses RFC 9110 (HTTP Semantics) which references RFC 3986 (URI Syntax). 



{\em Multiple Models:} While different LLMs may vary in performance on sub-tasks (e.g., constraint identification, triaging), we focus on pipeline-level design choices that are largely model-agnostic. Absolute performance can improve by choosing different models, but we expect the overall trends to remain consistent.


\section{Conclusion}
\label{sec:conclusion}

Extremal testing generates tests that occur near the boundaries of specifications. Our goal is to harden widely used Internet protocol implementations against messages with unexpected fields, or valid messages received in unexpected states which can degrade reliability, and in extreme cases expose dangerous vulnerabilities.



At a higher level, the main lesson of this paper is that LLMs are most useful here not as end-to-end test generators, but to convert natural-language specifications into explicit test targets. We use LLMs twice: first, to extract constraints from protocol specifications 
and only then systematically generate test cases at or near the boundaries of these constraints. 
Our approach is \emph{protocol-agnostic}, allowing users to specify their own test input and output formats as JSON.


We applied Extremal Testing to HTTP, DNS, BGP, SMTP, and the QUIC handshake, uncovering 42 bugs and inconsistencies. Of these, 26 have been acknowledged and 18 fixed by maintainers. Bugs span three broad categories: missing input validation (e.g., HTTP servers accepting malformed \texttt{Host} headers), state machine violations (e.g., SMTP servers accepting commands out of sequence), and semantic misinterpretation (e.g., GoBGP failing to reject AS loops in confederation paths). Our ablation study confirms that two-stage decomposition---constraint extraction followed by extremal test generation, applied section-by-section with cross-reference expansion---is essential, producing up to $22\times$ more anomalies than naive one-shot LLM generation.

Extremal testing's focus on identifying errors due to violations of constraints in RFCs makes it complementary to existing approaches to protocol implementation reliability, including fuzz testing, symbolic execution, and model-based testing. Together, they form a powerful ensemble of techniques that can make Internet protocol implementations more resilient and secure. Our work was supported in part by NSF Grant CNS-2402958.



\newpage
\bibliographystyle{plain}
\bibliography{paper}

\newpage
\appendix

\section{Test Format}
\label{app:test-format}

This appendix documents the input and output test formats used by each protocol-specific harness.

\subsection{HTTP}

\textbf{Test input format:} 

\begin{lstlisting}[style=jsonstyle]
{
    "test_id": "<integer>",
    "constraint": "<exact constraint that is being tested from the given list of constraints>",
    "description": "<description of the test case>",
    "tag": "<Constraint id from the given list of constraints. e.g. C1, C2, C3, ...>",
    "scheme": "<uri scheme>",
    "authority": "<uri authority>",
    "path": "<uri path>",
    "query": "<uri query>",
    "fragment": "<uri fragment>",
    "base_uri": "<base uri or null>",
    "expected_response_code": "<expected HTTP response code>",
    "expected_path": "<expected path after resolution>",
    "filesystem": {
        "<dir_path1>": ["<filename1>", "<filename2>", "..."],
        "<dir_path2>": ["<filename1>", "<filename2>", "..."],
        "<dir_pathN>": ["<filename1>", "<filename2>", "..."]
    },
    "symlinks": {
        "<symlink_path1>": "<target1>",
        "<symlink_path2>": "<target2>",
        "<symlink_pathN>": "<targetN>"
    }
}
\end{lstlisting}

\textbf{Test output format:} 

\begin{lstlisting}[style=jsonstyle]
{
    "status_code": "<HTTP status code (integer)>",
    "resolved_uri": "<resolved URI after any redirects or normalization (string)>"
}
\end{lstlisting}

\newpage
\subsection{DNS}
\textbf{Test input format:} 

\begin{lstlisting}[style=jsonstyle]
{
    "test_id": "<integer>",
    "constraint": "<exact constraint that is being tested from the given list of constraints>",
    "description": "<Brief description of the test case. Positive for barely valid input, Negative for barely invalid input>",
    "tag": "<Combination of Constraint id from the given list of constraints and Positive/Negative indicator. e.g. C1_positive, C2_negative, C3_positive, ...>",
    "zone":"campus.edu.\t500\tIN\tSOA\tns1.outside.edu. root.campus.edu. 8 6048 4000 2419200 6048\ncampus.edu.\t500\tIN\tNS\tns1.outside.edu.\na.a.a.test.\t500\tIN\tDNAME\tsome.domain.\n",
    "query": [{"Name": "*.a.test.", "Type": "DNAME"}]
}
\end{lstlisting}

\textbf{Test output format:} 

\begin{lstlisting}[style=jsonstyle]
{
    "Response": "response from the DNS server in a structured format"
}
\end{lstlisting}

\subsection{SMTP}

\textbf{Test input format:} 

\begin{lstlisting}[style=jsonstyle]
{
    "test_id": "<integer>",
    "constraint": "<exact constraint that is being tested from the given list of constraints>",
    "description": "<Brief description of the test case. Positive for barely valid input, Negative for barely invalid input>",
    "tag": "<Combination of Constraint id from the given list of constraints and Positive/Negative indicator. e.g. C1_positive, C2_negative, C3_positive, ...>",
    "prev_command_seq": ["<cmd1>", "<cmd2>", "..." ],
    "server_state": "<server state after executing the previous commands e.g. INIT, EHLO_RCVD, MAIL_FROM_RCVD, ... etc>",
    "command": "<SMTP command to be tested>",
    "expected_response": "<expected server response to the command>"
}
\end{lstlisting}

\textbf{Test output format:} 

\begin{lstlisting}[style=jsonstyle]
{
    "code": "<integer>, the SMTP response code from the server implementation"
}
\end{lstlisting}

\newpage

\subsection{BGP (Confederation)}

\textbf{Test input format:} 

\begin{lstlisting}[style=jsonstyle]
{
    "test_id": "<integer>",
    "constraint": "<exact constraint that is being tested from the given list of constraints>",
    "description": "<Brief description of the test case. Positive for barely valid input, Negative for barely invalid input>",
    "tag": "<Combination of Constraint id from the given list of constraints and Positive/Negative indicator. e.g. C1_positive, C2_negative, C3_positive, ...>",
    "originAS": "integer e.g. 512. This is the AS number of the originating router",
    "router2": {
        "asNumber": "integer e.g. 768. This is the AS number of the second router",
        "subAS": "integer e.g. 256. If this is non-zero, then it is part of a confederation and this is the sub-AS number within the confederation"
    },
    "router3": {
        "asNumber": "integer e.g. 1024. This is the AS number of the third router",
        "subAS": "integer e.g. 256. If this is non-zero, then it is part of a confederation and this is the sub-AS number within the confederation"
    },
    "removePrivateAS": "boolean. A flag for router 2. Similar to cisco remove-private-as command. If this is true, then the router 2 should remove all private AS numbers from its AS path before forwarding it to router 3.",
    "replaceAS": "boolean. A flag for router 2. If this is true, then the router should replace all private AS numbers with the confederation number before forwarding the route to router 3.",
    "localPref": "integer e.g. 50. This is the local preference value set by router 2 when advertising to router 3.",
    "isExternalPeer": "boolean. A flag for router 3. If this is True then in BGP config of router 3. the neighbor is configured as neighbor peer-as external. i.e. if this is enabled then connection will be denied if the AS number router 2 is same as mine."
}
\end{lstlisting}

\textbf{Test output format:} 

\begin{lstlisting}[style=jsonstyle]
{
    "isRIB2": "<boolean>, true if the route gets installed in RIB2",
    "aspath2": "<string>, AS path of the route at RIB2",
    "isRIB3": "<boolean>, true if the route gets installed in RIB3",
    "aspath3": "<string>, AS path of the route at RIB3"
}
\end{lstlisting}

\newpage

\subsection{QUIC}

\textbf{Test input format:} 

\begin{lstlisting}[style=jsonstyle]
{
    "test_id": "<integer>",
    "constraint": "<exact constraint that is being tested from the given list of constraints>",
    "description": "<description of the test case>",
    "tag": "<Constraint id from the given list of constraints. e.g. C1, C2, C3, ...>",
    "tested_party":"<client | server>",
    "mutations": [{
        "mutation": "<must be remove_field or modify_field>",
        "target": "<client | server>",
        "fields": {
            "field_name": "<name of the field to be removed or modified>",
            "new_value": "<new value for the field if mutation is modify_field, omit if mutation is remove_field>"
        }
    },
    {
        "<mutation>": "more mutations if needed"
    }],
    "expected_result": "<success | fail>, this describes the expected outcome of the tested party.",
    "_comment": "this is for your information only. do not generate this field. Possible values for field_name and their new_value types are the following. See next item in the array for example on how to generate bytes, etc. random: bytes, legacy_session_id: bytes, cipher_suites: list[int], legacy_compression_methods: list[int],alpn_protocols: list[str], early_data: bool, key_share: list[ tuple [int, bytes] ], psk_key_exchange_modes: Optional[list[int]], signature_algorithms: list[int], supported_versions: list[int] signature_algorithms: [list[int]], supported_groups: [list[int]] supported_versions: Optional[list[int]], other_extensions: list[tuple[int, bytes]]. "
}
\end{lstlisting}

\textbf{Test output format:} 

\begin{lstlisting}[style=jsonstyle]
{
    "handshake_status": "success / failure"
}
\end{lstlisting}

\section{LLM Prompts}
\label{app:prompts}
We use GPT 5 over the OpenAI API for the experiments.

\subsection{Constraint Generation Prompt}
\label{prompts:constraint_gen}
This prompt is used to extract testable constraints
from each section of the RFC.

System Prompt:
\begin{lstlisting}[style=jsonstyleplain]
You are an assistant that extracts 
*input-related constraints* from RFC for a
specific testing framework.

### Inputs:
- A test case format (in JSON) describing test 
  case fields
- A chunk of RFC text.

### Task:
1. Use the test case format to infer what
inputs can be controlled by the tests.
2. Scan the RFC chunk and find sentences that 
define constraints on those inputs.
   These include:
   - syntax rules,
   - allowed or disallowed values,
   - length or size limits,
   - character set restrictions,
   - relationships between multiple inputs,
   - ordering/state rules that can be
     represented as test inputs/state.
3. Constraints are generally RFC statements
that include MUST/MUST NOT/SHOULD/SHOULD NOT.
But also look for sentences that describe a 
rule or constraint on inputs that can be
tested with this framework.
4. Every constraint is written as a tuple: 
   (<section_number>, <constraint>).
5. If the chunk has no relevant constraints, 
return [].

### Important:
- Return each constraint sentence *exactly as
written* in the RFC (no edits).
- Only include sentences that can plausibly be
tested using the described setup.

### Output format (for each chunk):
Return ONLY a JSON array like:
[
  ["4.1.1", "sentence1"],
  ["4.1.1", "sentence2"],
  ["4.2",   "sentence3"]
]
No markdown, no explanation.

Here is the test case format you should assume
when deciding which RFC constraints are
testable: 

=== TEST CASE FORMAT START ===
< Test Case Format in JSON>
=== TEST CASE FORMAT END ===

Now, here is a section of the RFC. Extract
input-related constraints that can be tested 
with this test case format. Each constraint 
must be returned as a 2-element JSON array: 
["<section_number>", "<constraint_sentence>"].
Return ONLY the JSON array as specified in 
the system prompt.

=== RFC SECTION START ===
<RFC Section Text>
=== RFC SECTION END ===
\end{lstlisting}

\subsection{Test Generation Prompt}
\label{prompts:test_gen}
System Prompt:

\begin{lstlisting}[style=jsonstyleplain]
You are an assistant that generates 
*extremal test cases* from RFC constraints
for a specific testing framework.
\end{lstlisting}

This prompt is used to generate extremal tests
from each constraint identified earlier in the pipeline.

\begin{lstlisting}[style=jsonstyleplain]
### Inputs

You will receive:
- The test case format (JSON schema or 
  example object).
- A list of RFC constraints (sentences).

* A constraint is a sentence that describes
  a rule on inputs that the test framework
  can exercise (commands, arguments, states,
  etc.).

### Task

Your job is to generate **extremal tests** 
for these constraints.

Definition of extremal tests:
- Tests at the boundary conditions between 
  valid and invalid.
- "Almost valid": barely violates a constraint
  (e.g., one character too many, one invalid
  character).
- "Almost invalid": barely satisfies a
  constraint (e.g., minimum required length,
  edge of allowed range).
- Test the precise point where valid becomes
  invalid.
- Try to generate multiple extremal tests for
  each constraint, to cover all corner cases.
- Include both positive tests (barely valid)
  and negative tests (barely invalid).
- Consider interactions between components
  when relevant.
- Focus on generating tests that might result
  in crashes/divergent behavior of serious
  consequence for security and reliability.

Requirements for the output:
- Return ONLY a JSON array (no prose).
- Each element is one test case object that
  follows the test case format.
- Use ONLY the fields defined in the test
  format (no extra keys).
- Every test object MUST include a "constraint"
  field set to exactly one of the provided 
  constraint sentences (verbatim, nochanges).
- Every test object MUST include a "test_id"
  field (you may choose any unique string or
  number within this batch; uniqueness across
  batches will be handled later).

Be explicit and systematic in exploring 
boundary conditions, but keep the output
strictly as JSON.
\end{lstlisting}

The user prompt for this step includes more variables. As an example, we provide the full user prompt for \texttt{batch\_size = 2}, where CornerCase asks the LLM to generate tests for constraints C1, C2.

\begin{lstlisting}[style=jsonstyleplain]
Here is the test case format you must follow:

=== TEST CASE FORMAT START ===
<test input format>
=== TEST CASE FORMAT END ===

=== REFERENCED RFC SECTIONS (for context) ===
Section X: RFC section text
Section Y: RFC section text
...
=== END REFERENCED SECTIONS ===

Here is a batch of RFC constraints. Each line
has a section id, and the exact constraint 
sentence. Use these constraint sentences 
exactly (do NOT edit them) in the "constraint"
field of your tests. You must generate 
multiple extremal tests for each constraint, 
covering positive and negative edges.

###Constraints:
C1: [1.1] 
C2: [1.1] Constraint 2 text (references 
    Section X, Y)

Now generate extremal test cases for these
constraints.
- Output ONLY a JSON array of test objects.
- Each test object must:
  * Follow the test format fields.
  * Have a "constraint" field equal to exactly
    one of the sentences above.
  * Have a "test_id" field unique within this
    batch.
Do not output any explanation or text outside
the JSON array.
\end{lstlisting}

\subsection{Result Analysis Prompt}
\label{prompts:confidence}
This prompt lets the LLM assign a confidence score and analysis text for whether the output of selected tests show real RFC violations and bugs.

\textbf{System Prompt}:
\begin{lstlisting}[style=jsonstyleplain]
### Role
You are an assistant helping to triage
differences between {protocol_name}
implementations.

### Inputs
You will be given a batch of (or a single)
test results where multiple {protocol_name}
implementations produced different responses
for the same test case (zone + query).

### Description of the fields in test results
The test results for each test case are of
the form:

<Test Output Format in JSON>

### Task 
Your job for EACH test in the batch (or 
the single test) is to:
1. Reason about the differences between the
implementations' outputs, considering:
   - Is one or more implementation likely
     violating the RFC (a real bug)?
   - Could the difference be due to acceptable
     implementation-specific behavior?
   - Could the difference plausibly be explained
     or fixed by configuration (e.g., security
     settings, extensions enabled/disabled,
     strictness toggles)?
2. Write a short, clear comment that summarizes
your judgment, referencing:
   - which servers look suspicious,
   - whether this is likely a real bug vs.
   configuration/behavioral choice,
   - any caveats or uncertainty you have.
3. Assign a confidence score from 0 to 10 that
represents how confident you are that
   this represents a REAL BUG (i.e., at least one
   implementation is non-compliant with the RFC):
   - 0 = very likely NOT a bug (probably
     config/expected behavior),
   - 10 = almost certainly a real bug.

### Output Format
Output format for each test:
- You MUST output an array of objects, one
  object per test, of the form:

  {{
    "test_id": <same integer test_id as input>,
    "comment": "<your explanation>",
    "confidence": <integer 0-10>
  }}

Constraints:
- Return ONLY a JSON array (no prose,
  no markdown).
- Do NOT omit any test from the batch; every
  input test must have exactly one output object.
- Do NOT invent test_ids; they must match the
  ones you received.
\end{lstlisting}

\textbf{User Prompt}

\begin{lstlisting}[style=jsonstyleplain]
### Test Results Batch (batch size {batch_size}):

Here is a batch of test results where different
implementations returned different responses.
For EACH test in this batch, you must output
an analysis object as described in the system
prompt.

### Tests
<One or more Test Output JSON>

### Start Analysis:  
Now, for this batch, return ONLY a JSON array
of analysis objects of the form:

{ "test_id": <int>, 
"comment": "<text>",
"confidence": <int 0-10> }.

Make sure every test above has exactly one
corresponding analysis object.
\end{lstlisting}

\section{Implementation Matrix}
\label{app:impl-matrix}

\begin{table*}[!htbp]
\centering
\fontsize{8}{9}\selectfont
\setlength{\tabcolsep}{4pt}
\begin{tabular}{p{1.9cm}p{2.2cm}p{1.2cm}p{8.8cm}}
\toprule
\textbf{Protocol} & \textbf{Implementation} & \textbf{Lang.} & \textbf{Description} \\
\midrule

DNS & BIND & C & De facto standard DNS implementation, widely deployed in production. \\
DNS & NSD & C & Authoritative-only DNS server, commonly used by TLD and ccTLD operators. \\
DNS & Knot DNS & C & High-performance authoritative DNS server with modern DNS feature support. \\
DNS & PowerDNS & C++ & Widely used DNS platform with flexible authoritative deployment support. \\
DNS & CoreDNS & Go & Cloud-native DNS server widely used in Kubernetes deployments. \\
DNS & YADIFA & C & Authoritative DNS server developed in the EURid ecosystem. \\
DNS & HickoryDNS & Rust & Rust-based DNS stack emphasizing safety and modern implementation design. \\
DNS & gdnsd & C & Lightweight authoritative DNS daemon designed for high-throughput serving. \\
DNS & TwistedNames & Python & Python-based DNS server from the Twisted ecosystem, useful as a contrasting implementation style. \\
DNS & Technitium & C\# & Feature-rich DNS server used in self-hosted and enterprise settings. \\

\midrule

HTTP & nginx & C & High-performance web server and reverse proxy, widely deployed in production. \\
HTTP & Apache httpd & C & Long-standing modular HTTP server with broad compatibility and deployment footprint. \\
HTTP & Caddy & Go & Modern HTTP server known for automatic HTTPS and simple configuration. \\
HTTP & H2O & C & Performance-oriented HTTP server with strong HTTP/2 support. \\
HTTP & lighttpd & C & Lightweight HTTP server often used in resource-constrained environments. \\

\midrule

SMTP & smtpd & Python & Python standard-library SMTP server, serving as a simple baseline implementation. \\
SMTP & aiosmtpd & Python & Asyncio-based SMTP server framework representing modern Python async behavior. \\
SMTP & OpenSMTPD & C & Security-focused mail transfer agent with relatively strict SMTP semantics. \\
SMTP & Mailpit & Go & Lightweight SMTP testing server commonly used in development environments. \\
SMTP & Stalwart & Rust & Modern Rust mail server emphasizing safety and integrated mail functionality. \\

\midrule

BGP(Confed.) & FRR & C & Mature open-source routing suite widely used in operational BGP deployments. \\
BGP(Confed.) & GoBGP & Go & API-friendly BGP implementation suited to programmable control-plane experimentation. \\
BGP(Confed.) & Batfish & Java & Control-plane analysis engine used to model and validate routing behavior. \\

\midrule

QUIC & quic-go & Go & Widely used Go QUIC stack and a common interoperability baseline. \\
QUIC & go-x-net & Go & QUIC implementation from the Go networking ecosystem used for comparison. \\
QUIC & picoquic & C & Lightweight C QUIC implementation often used in experimentation and interop testing. \\
QUIC & HAProxy & C & Deployment-oriented QUIC-capable implementation integrated into a production load balancer. \\
QUIC & MsQuic & C & Microsoft's high-performance cross-platform QUIC implementation. \\
QUIC & mvfst & C++ & Meta's QUIC stack, designed for large-scale production deployment. \\
QUIC & nginx & C & QUIC-enabled NGINX implementation integrated with mainstream web-serving workflows. \\
QUIC & quinn & Rust & Rust QUIC implementation emphasizing safety and clean transport abstractions. \\
QUIC & neqo & Rust & Mozilla's QUIC/TLS stack used in browser and protocol experimentation. \\
QUIC & quiche & Rust & Cloudflare's QUIC implementation with broad visibility in the ecosystem. \\
QUIC & lsquic & C & LiteSpeed's C QUIC implementation optimized for practical deployment. \\
QUIC & aioquic & Python & Python QUIC implementation that we modified to support handshake-field mutation. \\
QUIC & ngtcp2 & C & Standards-focused C QUIC implementation with strong conformance emphasis. \\
QUIC & kwik & Java & Java QUIC implementation providing JVM-based interoperability coverage. \\

\bottomrule
\end{tabular}
\caption{Protocol-wise implementation matrix used in our evaluation, including implementation language and a brief distinguishing characteristic of each system.}
\label{tab:impl-matrix}
\end{table*}

\end{document}